\documentclass[a4paper]{article}
\usepackage{graphicx}
\usepackage{amsmath}
\usepackage{babel}

\input{tcilatex}

\begin{document}

\author{S. Manoff \\
\textit{Bulgarian Academy of Sciences}\\
\textit{\ Institute for Nuclear Research and Nuclear Energy}\\
\textit{\ Department of Theoretical Physics}\\
\textit{\ Blvd. Tzarigradsko Chaussee 72}\\
\textit{\ 1784 Sofia - Bulgaria}}
\title{Conformal derivative and conformal transports over $(L_n,g)$ spaces}
\date{\textit{e-mail address: smanov@inrne.bas.bg}}
\maketitle

\begin{abstract}
Transports preserving the angle between two contravariant vector fields but
changing their lengths proportional to their own lengths are introduced as
''conformal'' transports and investigated over $(L_n,g)$-spaces. They are
more general than the Fermi-Walker transports. In an analogous way as in the
case of Fermi-Walker transports a conformal covariant differential operator
and its conformal derivative are defined and considered over $(L_n,g)$%
-spaces. Different special types of conformal transports are determined
inducing also Fermi-Walker transports for orthogonal vector fields as
special cases. Conditions under which the length of a non-null contravariant
vector field could swing as a homogeneous harmonic oscillator are
established. The results obtained regardless of any concrete field
(gravitational) theory could have direct applications in such types of
theories.

\end{abstract}

\section{Introduction}

The construction of a frame of reference for an accelerated observer by
means of vector fields preserving their lengths and the angles between them
under a Fermi-Walker transport \cite{Manoff-2}, \cite{Manoff-3}, could also
be related to the description of the motion of the axes of a gyroscope in a
space with an affine connection and metrics [$(L_n,g)$-space]. On the other
side, the problem arises how can we describe the motion of vector fields
preserving the angles between them but, at the same time, changing the
length of every one of them proportionally to its own length. In special
cases of $(L_n,g)$-spaces there are different solutions of this problem
which induces the definition of a conformal transport.

\textbf{Definition 1}. The \textit{conformal transport} is a special type of
transport (along a contravariant vector field) under which the change of the
length of a contravariant vector field is proportional to the length itself
and the angle between two contravariant vector fields does not change.

\textit{Remark}. The most notions, abbreviations, and symbols in this paper
are defined in the previous papers \cite{Manoff-2}, \cite{Manoff-3}. The
reader is kindly asked to refer to one of them.

\subsection{Conformal transports in $M_n$-, $V_n$-, and $U_n$-spaces}

(a) In flat and (pseudo) Riemannian spaces without or with torsion ($M_n$-, $%
V_n$-, and $U_n$- spaces respectively, $\dim M_n=n$), [$\nabla _ug=0$ for $%
\forall u\in T(M)$] the mentioned above problem could be easily solved by
means of a conformal (angle preserving) mapping leading to the notion of 
\textit{metric conformal to a given metric} \cite{Raschewski}.

If we construct the length of $\xi $ by the use of the metric $g$ in a (for
instance) $V_n$-space as $l_\xi =\,\mid g(\xi ,\xi )\mid ^{\frac 12}$ and by
the use of the conformal to $g$ metric $\widetilde{g}=e^{2\varphi }.g$, $%
\varphi =\varphi (x^k)\in C^r(M)$, $r\geq 1$, as $\widetilde{l}_\xi =\,\mid 
\widetilde{g}(\xi ,\xi )\mid ^{\frac 12}=e^\varphi .l_\xi $, then the rate
of change $u\widetilde{l}_\xi $ of the length of $\xi $ along a
contravariant vector field $u$ leads to the relation $u\widetilde{l}_\xi
=(u\varphi ).\widetilde{l}_\xi +e^\varphi .ul_\xi $. If $ul_\xi =0$, then $u%
\widetilde{l}_\xi =(u\varphi ).\widetilde{l}_\xi $ and therefore, the change 
$u\widetilde{l}_\xi $ of the length $\widetilde{l}_\xi $ is proportional to $%
\widetilde{l}_\xi $. This is the case when $\xi $ fulfils the equation $%
\nabla _u\xi =0$ determining a parallel transport of $\xi $ along $u$
because of the relation in a $M_n$-, $V_n$-, or $U_n$-space

\begin{equation}  \label{1.1}
ul_\xi =\pm \frac 1{l_\xi }.g(\nabla _u\xi ,\xi )\text{ , \thinspace
\thinspace \thinspace \thinspace }l_\xi \neq 0\text{ .}
\end{equation}

The change of the cosine of the angle $[\cos (\xi ,\eta )=(l_\xi .l_\eta
)^{-1}.g(\xi ,\eta )]$ between two contravariant vector fields $\xi $ and $%
\eta $ is described in these types of spaces by the relation 
\begin{equation*}
u[\cos (\xi ,\eta )]=\frac 1{l_\xi .l_\eta }.[g(\nabla _u\xi ,\eta )+g(\xi
,\nabla _u\eta )]- 
\end{equation*}
\begin{equation}  \label{1.2}
-[\frac 1{l_\xi }.(ul_\xi )+\frac 1{l_\eta }.(ul_\eta )].\cos (\xi ,\eta )%
\text{ , \thinspace \thinspace \thinspace \thinspace \thinspace \thinspace
\thinspace \thinspace \thinspace }l_\xi \neq 0\text{ , \thinspace \thinspace
\thinspace \thinspace }l_\eta \neq 0\text{ .}
\end{equation}

Under the conditions for parallel transport ($\nabla _u\xi =0$, $\nabla
_u\eta =0$) of $\xi $ and $\eta $, it follows that $u[\cos (\xi ,\eta )]=0$.
Since the cosine between $\xi $ and $\eta $ defined by the use of the
conformal to $g$ metric $\widetilde{g}$ as 
\begin{equation*}
\cos (\xi ,\eta )=\frac 1{\widetilde{l}_\xi .\widetilde{l}_\eta }.\widetilde{%
g}(\xi ,\eta )=\frac{e^{2\varphi }.g_{ij}.\xi ^i.\eta ^j}{e^{2\varphi }.\mid
g_{kl}.\xi ^k.\xi ^l\mid ^{\frac 12}.\mid g_{mn}.\eta ^m.\eta ^n\mid ^{\frac
12}}= 
\end{equation*}

\begin{equation}  \label{1.3}
=\frac{g_{ij}.\xi ^i.\eta ^j}{\mid g_{kl}.\xi ^k.\xi ^l\mid ^{\frac 12}.\mid
g_{mn}.\eta ^m.\eta ^n\mid ^{\frac 12}}=\frac 1{l_\xi .l_\eta }.g(\xi ,\eta )
\end{equation}

\noindent does not change by the replacement of the metric $g$ with the
metric $\widetilde{g}$, it follows that $u[\cos (\xi ,\eta )]=0$ for $u%
\widetilde{l}_\xi =(u\varphi ).\widetilde{l}_\xi $, and $u\widetilde{l}_\eta
=(u\varphi ).\widetilde{l}_\eta $. This means that \textit{parallel
transports in a given }$M_n$\textit{-, }$V_n$\textit{-, or }$U_n$\textit{%
-space induce conformal transports in the corresponding conformal space}.

If $ul_\xi =0$, $ul_\eta =0$ are not valid, then the proportionality of $u%
\widetilde{l}_\xi $ to $\widetilde{l}_\xi $ and of $u\widetilde{l}_\eta $ to 
$\widetilde{l}_\eta $ is violated even in these spaces. This fact induces
the problem of finding out transports different from the parallel transport (%
$\nabla _u\xi =0$, $\nabla _u\eta =0$) under which the angle between two
contravariant vector fields does not change and at the same time the rate of
change of the lengths of these vector fields is proportional to the
corresponding length.

(b) In Weyl's spaces ($W_n$-spaces) [$\nabla _ug=\frac 1n.Q_u.g$ for $%
\forall u\in T(M)$], it follows from the general relations for $ul_\xi $ and 
$u[\cos (\xi ,\eta )]$ in $(L_n,g)$-spaces 
\begin{equation}  \label{1.4}
ul_\xi =\pm \frac 1{2.l_\xi }.[(\nabla _ug)(\xi ,\xi )+2.g(\nabla _u\xi ,\xi
)]\text{ , \thinspace \thinspace \thinspace \thinspace \thinspace \thinspace
\thinspace \thinspace \thinspace }l_\xi \neq 0\text{ ,}
\end{equation}

\begin{equation*}
u[\cos (\xi ,\eta )]=\frac 1{l_\xi .l_\eta }.[(\nabla _ug)(\xi ,\eta
)+g(\nabla _u\xi ,\eta )+g(\xi ,\nabla _u\eta )]- 
\end{equation*}
\begin{equation}  \label{1.5}
-[\frac 1{l_\xi }.(ul_\xi )+\frac 1{l_\eta }.(ul_\eta )].\cos (\xi ,\eta )%
\text{ , \thinspace \thinspace \thinspace \thinspace \thinspace \thinspace
\thinspace \thinspace \thinspace }l_\xi \neq 0\text{ , \thinspace \thinspace
\thinspace \thinspace }l_\eta \neq 0\text{ ,}
\end{equation}

\noindent that under the conditions for parallel transport ($\nabla _u\xi =0$%
, $\nabla _u\eta =0$) of $\xi $ and $\eta $ 
\begin{equation}
ul_\xi =\frac 1{2.n}.Q_u.l_\xi \text{ , \thinspace \thinspace \thinspace
\thinspace \thinspace \thinspace \thinspace \thinspace \thinspace \thinspace
\thinspace }ul_\eta =\frac 1{2.n}.Q_u.l_\eta \text{ ,}  \label{1.6}
\end{equation}

\begin{equation}  \label{1.7}
u[\cos (\xi ,\eta )]=0
\end{equation}

Therefore, the \textit{parallel transports in }$W_n$\textit{-spaces are at
the same time conformal transports}.

(c) The case of $(L_n,g)$-spaces with equal to zero trace-free part of the
covariant derivative of the metric $g$ is analogous to this of Weyl's spaces 
$W_n$.

\subsection{Extended covariant differential operator}

In previous papers \cite{Manoff-2}, \cite{Manoff-3}, we have considered
Fermi-Walker transports over $(L_n,g)$ and $(\overline{L}_n,g)$-spaces
leading to preservation of the lengths of two contravariant vector fields
and the angle between them, when they are transported along a non-null
contravariant vector field. The investigations have been based on a special
form of the extended covariant differential operator $^e\nabla _u$
determined over a $(L_n,g)$- or a $(\overline{L}_n,g)$-space as $^e\nabla
_u=\nabla _u-\overline{A}_u$, where $\nabla _u$ is the covariant
differential operator: 
\begin{equation*}
\nabla _u:v\rightarrow \nabla _uv=\overline{v}\text{ , \thinspace \thinspace
\thinspace \thinspace \thinspace \thinspace \thinspace \thinspace \thinspace
\thinspace \thinspace \thinspace }v,\,\,\overline{v}\in \otimes ^k\,_l(M)%
\text{ .} 
\end{equation*}

The vector field $u$ is a contravariant vector field, $u\in T(M)$, $v$ and $%
\overline{v}$ are tensor fields with given contravariant rank $k$ and
covariant rank $l$.

$\overline{A}_u$ appears as a mixed tensor field of second rank but acting
on tensor fields as a covariant differential operator because $^e\nabla _u$
is defined as covariant differential operator with the same properties as
the covariant differential operator $\nabla _u$. In fact, $\overline{A}_u$
can be defined as $\overline{A}_u=\nabla _u-\,^e\nabla _u$. If $\overline{A}%
_u$ is a given mixed tensor field, then $^e\nabla _u$ can be constructed in
a predetermined way.

In accordance to its property \cite{Manoff-2} $\overline{A}_{u\,+\,\,v}=%
\overline{A}_u+\overline{A}_v$, $\overline{A}_u$ has to be linear to $u$.
The existence of covariant and contravariant metrics $g$ and $\overline{g}$
in a $(L_n,g)$-space allows us to represent $\overline{A}_u$ in the form $%
\overline{A}_u=\overline{g}(A_u)$. There are at least three possibilities
for construction of a covariant tensor field of second rank $A_u$ in such a
way that $A_u$ is linear to $u$, i. e. we can determine $A_u$ as

A. $A_u=C(u)=A(u)=A_{ijk}.u^k.dx^i\otimes dx^j$ .

B. $A_u=C(u)=\nabla _uB=B_{ij;k}.u^k.dx^i\otimes dx^j$ .

C. $A_u=C(u)=A(u)+\nabla _uB=(A_{ijk}+B_{ij;k}).u^k.dx^i\otimes dx^j$.

These three possibilities for definition of $A_u$ lead to three types ($A$, $%
B$, and $C$ respectively) of the extended covariant differential operator $%
^e\nabla _u=\nabla _u-\overline{g}(A_u)$. $A_u$ can also obey additional
conditions determining the structure of the mixed tensor field $\overline{A}%
_u=\overline{g}(A_u)$. One can impose given conditions on $^e\nabla _u$
leading to predetermined properties of $\overline{A}_u$ and vice versa: one
can impose conditions on the tensor field $\overline{A}_u$ leading to
predetermined properties of $^e\nabla _u$. The method of finding out the
conditions for Fermi-Walker transports over $(L_n,g)$-spaces allow us to
consider other types of transports with important properties for describing
the motion of physical systems over such type of spaces.

\subsection{Problems and results}

In Sec. 2 conformal transports over $(L_n,g)$-spaces are determined and
considered with respect to their structure. A conformal covariant
differential operator and its corresponding conformal derivative are
introduced. In Sec. 3 Fermi-Walker transports along a contravariant vector
field for orthogonal to it contravariant vector fields are found on the
basis of the structure of conformal transports. In Sec. 4 conformal
transports of null vector fields are discussed. In Sec. 5 the length of a
contravariant vector field as homogeneous harmonic oscillator with given
frequency is considered on the grounds of a conformal transport. Sec. 6
comprises some concluding remarks.

\textit{Remark}. All formulas written in index-free form are identical and
valid in their form (but not in their contents) for $(L_n,g)$- and $(%
\overline{L}_n,g)$-spaces. The difference between them appears only if they
are written in a given (co-ordinate or non-co-ordinate) basis. The conformal
derivative and the conformal transports over $(\overline{L}_n,g)$-spaces
will be considered elsewhere\cite{Manoff-4}.

\section{Conformal transports over $(L_n,g)$-spaces}

Let us now take a closer look at the notion of conformal transport over $%
(L_n,g)$-spaces.

\textit{Remark}. The following below conditions are introduced on the
analogy of the case of Fermi-Walker transports.

The main assumption related to the notion of conformal transport and leading
to a definition of an external covariant differential operator $^e\nabla
_u=\,^c\nabla _u$ (called conformal covariant differential operator) is that
the parallel transports $^c\nabla _u\xi =0$ and $^c\nabla _u\eta =0$ of two
contravariant non-null vector fields $\xi $ and $\eta $ respectively induce
proportional to the lengths $l_\xi =\,\mid g(\xi ,\xi )\mid ^{\frac 12}$ and 
$l_\eta =\,\mid g(\eta ,\eta )\mid ^{\frac 12}$ changes of $l_\xi $ and $%
l_\eta $ along a contravariant vector field $u$ as well as preservation of
the angle between $\xi $ and $\eta $ with respect to a special transport of
the covariant differential operator $\nabla _u$. The rate of change of the
length $l_\xi $ of a non--null contravariant vector field $\xi $ and the
rate of change of the cosine of the angle between two non-null contravariant
vector fields $\xi $ and $\eta $ over a $(L_n,g)$-space can be found in the
forms (\ref{1.4}) and (\ref{1.5}). If $^c\nabla _u\xi =\nabla _u\xi -%
\overline{A}_u\xi =0$ and $^c\nabla _u\eta =\nabla _u\eta -\overline{A}%
_u\eta =0$, then the conditions have to be fulfilled 
\begin{equation}  \label{2.1}
ul_\xi =\pm \frac 1{2.l_\xi }.[(\nabla _ug)(\xi ,\xi )+2.g(\overline{A}_u\xi
,\xi )]=f(u).l_\xi \text{ , \thinspace \thinspace \thinspace \thinspace
\thinspace \thinspace \thinspace }l_\xi \neq 0\text{ , \thinspace \thinspace
\thinspace \thinspace \thinspace }f(u)\in C^r(M)\text{ ,}
\end{equation}
\begin{equation*}
u[\cos (\xi ,\eta )]=\frac 1{l_\xi .l_\eta }.[(\nabla _ug)(\xi ,\eta )+g(%
\overline{A}_u\xi ,\eta )+g(\xi ,\overline{A}_u\eta )]- 
\end{equation*}
\begin{equation}  \label{2.2}
-[\frac 1{l_\xi }.(ul_\xi )+\frac 1{l_\eta }.(ul_\eta )].\cos (\xi ,\eta )=0%
\text{ .}
\end{equation}

From the first condition, under the assumption $\overline{A}_u=\overline{g}%
(A_u)$ with $A_u=C(u)=C_{ij}(u).dx^i\otimes dx^j$, we obtain 
\begin{equation}  \label{2.3}
ul_\xi =\pm \frac 1{2.l_\xi }.[(\nabla _ug)(\xi ,\xi )+2.g(\overline{g}%
(C(u))(\xi ),\xi )]=f(u).l_\xi \text{ ,}
\end{equation}

where \cite{Manoff-2} 
\begin{equation}  \label{2.4}
\begin{array}{c}
g( \overline{g}(C(u))(\xi ),\xi )=[_sC(u)](\xi ,\xi )=\,_sC_{jk}(u).\xi
^j.\xi ^k\text{ ,} \\ 
_sC(u)=C_{(jk)}(u).dx^j.dx^k \text{ , \thinspace \thinspace \thinspace }%
C_{(jk)}(u)=\frac 12.[C_{jk}(u)+C_{kj}(u)] \\ 
\text{\thinspace \thinspace }dx^j.dx^k=\frac 12(dx^j\otimes dx^k+dx^k\otimes
dx^j)\text{ .}
\end{array}
\end{equation}

On the other hand, $(\nabla _ug)(\xi ,\xi )=g_{jk;m}.u^m.\xi ^j.\xi ^k$.
Therefore, the condition $ul_\xi =f(u).l_\xi $ for $\forall \xi \in T(M)$
leads to the relations 
\begin{equation*}
ul_\xi =\pm \frac 1{2.l_\xi }.\{(\nabla _ug)(\xi ,\xi )+2.[_sC(u)](\xi ,\xi
)\}=f(u).l_\xi \text{ , \thinspace \thinspace \thinspace \thinspace
\thinspace \thinspace \thinspace }l_\xi \neq 0\text{ , \thinspace \thinspace
\thinspace \thinspace \thinspace }f(u)\in C^r(M)\text{ ,} 
\end{equation*}
\begin{equation}  \label{2.5}
\begin{array}{c}
(\nabla _ug)(\xi ,\xi )+2.[_sC(u)](\xi ,\xi )=\pm 2.f(u).l_\xi
^2=2.f(u).g(\xi ,\xi ) \text{ for }\forall \xi \in T(M)\text{ ,} \\ 
\lbrack \nabla _ug+2._sC(u)-2.f(u).g](\xi ,\xi )=0 \text{ for }\forall \xi
\in T(M)\text{ ,} \\ 
\nabla _ug=-2._sC(u)+2.f(u).g:\,_sC(u)=-\frac 12.\nabla _ug+f(u).g\text{ .}
\end{array}
\end{equation}

Since $C(u)$ and respectively $_sC(u)$ have to be linear to $u$, $f(u)$
should have the form $f(u)=f_k.u^k$,\thinspace \thinspace \thinspace $f\in
T^{*}(M)$. Therefore, 
\begin{equation}  \label{2.6}
_sC(u)=-\,\frac 12.\nabla _ug+f(u).g\text{ , \thinspace \thinspace
\thinspace \thinspace \thinspace \thinspace }u\in T(M)\text{ , \thinspace
\thinspace \thinspace \thinspace }f\in T^{*}(M)\text{ , for }\forall \xi \in
T(M)\text{ .}
\end{equation}

By the use of the condition $ul_\xi =f(u).l_\xi $ for $\forall \xi \in T(M)$
and $l_\xi \neq 0$, we have found the explicit form of the symmetric part $%
_sC(u)$ of $C(u)=\,_sC(u)+\,_aC(u)$, where 
\begin{equation}  \label{2.7}
\begin{array}{c}
_aC(u)=C_{[jk]}(u).dx^j\wedge dx^k \text{ , \thinspace \thinspace }%
C_{[jk]}(u)=\frac 12.[C_{jk}(u)-C_{kj}(u)]\text{ ,} \\ 
dx^j\wedge dx^k=\frac 12(dx^j\otimes dx^k-dx^k\otimes dx^j)\text{ .}
\end{array}
\end{equation}

If we now assume the validity of the first condition $ul_\xi =f(u).l_\xi $
[fulfilled for $_sC(u)=-\,\frac 12.\nabla _ug+f(u).g$], then from the
expression for $u[\cos (\xi ,\eta )]$ we obtain the relations: 
\begin{equation*}
u[\cos (\xi ,\eta )]=\frac 1{l_\xi .l_\eta }.[(\nabla _ug)(\xi ,\eta )+g(%
\overline{A}_u\xi ,\eta )+g(\xi ,\overline{A}_u\eta )]- 
\end{equation*}
\begin{equation}  \label{2.8}
-[\frac 1{l_\xi }.f(u).l_\xi +\frac 1{l_\eta }.f(u).l_\eta ].\cos (\xi ,\eta
)\text{ , \thinspace \thinspace \thinspace for }\forall \xi ,\eta \in T(M)%
\text{ ,}
\end{equation}
\begin{equation}  \label{2.9}
\begin{array}{c}
g( \overline{A}_u\xi ,\eta )=g(\overline{g}(C(u))(\xi ),\eta
)=g_{ij}.g^{il}.C_{lk}(u).\xi ^k.\eta ^j=\,C_{jk}(u).\eta ^j.\xi
^k=[C(u)](\eta ,\xi )\text{ ,} \\ 
g( \overline{A}_u\eta ,\xi )=g(\overline{g}(C(u))(\eta ),\xi )=[C(u)](\xi
,\eta )\text{ ,} \\ 
(\nabla _ug)(\xi ,\eta )+g( \overline{A}_u\xi ,\eta )+g(\xi ,\overline{A}%
_u\eta )=(\nabla _ug)(\xi ,\eta )+[_sC(u)](\eta ,\xi )+[_aC(u)](\eta ,\xi )+
\\ 
+[_sC(u)](\xi ,\eta )+[_aC(u)](\xi ,\eta )\text{ .}
\end{array}
\end{equation}

Since 
\begin{equation*}
\begin{array}{c}
\lbrack _sC(u)](\eta ,\xi )=[_sC(u)](\xi ,\eta )=-\,\frac 12.(\nabla
_ug)(\xi ,\eta )+f(u).g(\xi ,\eta ) \text{ } \\ 
\text{ and }[_aC(u)](\eta ,\xi )=-[_aC(u)](\xi ,\eta )\text{ ,}
\end{array}
\end{equation*}

\noindent it follows for $u[\cos (\xi ,\eta )]$%
\begin{equation}
u[\cos (\xi ,\eta )]=0\text{ ,\thinspace \thinspace \thinspace \thinspace
\thinspace \thinspace \thinspace \thinspace \thinspace \thinspace \thinspace
\thinspace \thinspace \thinspace \thinspace \thinspace \thinspace \thinspace
\thinspace \thinspace \thinspace \thinspace }g(\xi ,\eta )=l_\xi .l_\eta
.\cos (\xi ,\eta )\text{ .}  \label{2.10}
\end{equation}

Therefore, $C(u)$ will have the explicit form 
\begin{equation}  \label{2.11}
C(u)=\,_aC(u)-\frac 12.\nabla _ug+f(u).g\text{ .}
\end{equation}

Now, we can define the notion of conformal covariant differential operator.

\textbf{Definition 2}. A \textit{conformal covariant differential operator} $%
^c\nabla _u$ in a $(L_n,g)$-space. An extended covariant operator $^e\nabla
_u$ with the structure 
\begin{equation*}
^e\nabla _u=\,^c\nabla _u=\nabla _u-\overline{g}(C(u))\text{ ,} 
\end{equation*}

\noindent where 
\begin{equation*}
C(u)=_aC(u)-\frac 12.\nabla _ug+f(u).g\text{ , \thinspace \thinspace } 
\end{equation*}
\begin{equation*}
\text{\thinspace }C(u)\in \otimes _2(M)\text{, \thinspace \thinspace
\thinspace \thinspace \thinspace \thinspace \thinspace }_aC(u)\in \Lambda
^2(M)\text{ ,\thinspace \thinspace \thinspace \thinspace \thinspace
\thinspace }u\in T(M)\text{ ,\thinspace \thinspace \thinspace \thinspace
\thinspace \thinspace }f\in T^{*}(M)\text{ , \thinspace \thinspace
\thinspace }g\otimes _{s2}(M)\text{ .} 
\end{equation*}

\noindent is called conformal covariant differential operator. It is denoted
as $^c\nabla _u$.

\textit{Remark}. For $f(u)=0:\,^{\,c}\nabla _u=\,^F\nabla _u$, i. e. for $%
f(u)=0 $ the conformal covariant differential operator $^c\nabla _u$ is
identical with the Fermi covariant differential operator $^F\nabla _u$.

On the analogy of the case of $^F\nabla _u$ we can have three types $%
(A,\,B,\,$and $C)$ of $^c\nabla _u$:

\small%
%
\textbf{Table 1}. Types of conformal covariant differential operator $%
^c\nabla _u$

\begin{center}
\fbox{$
\begin{array}{ll}
\underline{\text{Type of }^c\nabla _u} & \underline{\text{Form of }C(u)} \\ 
A & C(u)=A(u)=A_{ijk}.u^k.dx^i\otimes dx^j \\ 
B & C(u)=\nabla _uB=B_{ij;k}.u^k.dx^i\otimes dx^j \\ 
C & C(u)=A(u)+\nabla _uB=(A_{ijk}+B_{ij;k}).u^k.dx^i\otimes dx^j
\end{array}
$}
\end{center}

\normalsize%
%

The first two types $A$ and $B$ are special cases of type $C$.

From the explicit form of $C(u)=\,_aC(u)-\frac 12\nabla _ug+f(u).g$ we can
choose $_aC(u)=\,_aA(u)$ and $_sC(u)=\,_sA(u)+\nabla _uB$ with $%
_sA(u)=f(u).g $ and $B=-\frac 12.g$.

A type of general ansatz (without too big loss of generality) for $%
A(u)=\,_aA(u)+\,_sA(u)=\,_aA(u)+f(u).g$ (keeping in mind its linearity to $u$
and the form of $\overline{A}_u=u^k.\overline{A}_{\partial _k}$) has the
form \cite{Manoff-2} 
\begin{equation}  \label{2.12}
A_{ij}(u)=p_k.u^k.^F\omega _{ij}+f_k.u^k.g_{ij}\text{ ,}
\end{equation}

\noindent or 
\begin{equation}
A(u)=p(u).^F\omega +f(u).g\text{ ,}  \label{2.13}
\end{equation}

\noindent where 
\begin{equation*}
^F\omega _{ij}=-\,^F\omega _{ji}\text{ , \thinspace \thinspace \thinspace }%
_aA(u)=\frac 12.[A_{ij}(u)-A_{ji}(u)].dx^i\wedge dx^j\text{ ,} 
\end{equation*}

$p,f\in T^{*}(M)$ are arbitrary given covariant vector fields, $%
p(u)=p_k.u^k=C(p,u)$,\thinspace \thinspace \thinspace $f(u)=f_k.u^k=C(f,u)$, 
$C(\partial _j,dx^i)=g_j^i=\delta _j^i$, $^F\omega =\,^F\omega
_{ij}.dx^i\wedge dx^j$ is an arbitrary given covariant antisymmetric tensor
field of second rank. Therefore, $\overline{g}(A(u))=p(u).\overline{g}%
(^F\omega )+f(u).\overline{g}(g)$,\thinspace \thinspace \thinspace $%
\overline{g}(g)=g_k^i.\partial _i\otimes dx^k$. If we express $p$ and $f$ by
the use of their corresponding with respect to the metric $g$ contravariant
vector fields $b=\overline{g}(p):g(b)=p$, and $q=\overline{g}(f):g(q)=f$
respectively, then $\overline{A}_u$ will obtain the form 
\begin{equation}
\overline{A}_u=\overline{g}(C(u))=g(b,u).\overline{g}(^F\omega )+g(q,u).%
\overline{g}(g)-\frac 12.\overline{g}(\nabla _ug)\text{ , \thinspace
\thinspace \thinspace \thinspace \thinspace \thinspace }b,q\in T(M)\text{ .}
\label{2.14}
\end{equation}

We have now the free choice of the contravariant vector fields $b$ and $q$,
which could depend on the physical problem to be considered. For the
determination of Fermi-Walker transports the vector field $b$ has been
chosen as $b=\frac 1e.u$ with $e=g(u,u)\neq 0$. There are other
possibilities for the choice of $b$ and $q$.

The conformal covariant differential operator will now have the form 
\begin{equation}  \label{2.15}
^c\nabla _u=\nabla _u-\overline{A}_u=\nabla _u-[g(b,u).\overline{g}(^F\omega
)+g(q,u).\overline{g}(g)-\frac 12.\overline{g}(\nabla _ug)]\text{ .}
\end{equation}

The result $^c\nabla _uv$ of the action of a conformal covariant
differential operator $^c\nabla _u$ (of type $A$, $B$, and $C$) on a tensor
field $v\in \otimes ^k\,_l(M)$ is called \textit{conformal derivative} of
type $A$, $B$, and $C$ respectively of the tensor field $v$.

\section{Fermi-Walker transports for orthogonal to $u$ vector fields}

The free choice of the vector fields $b$ and $q$ allows us to determine
another type of Fermi-Walker transport for orthogonal to $u$ vector fields
than the defined in \cite{Manoff-2}.

(a) If we chose $b=\frac 1e.u$ and $q=\xi $ in the expression for $^c\nabla
_u\xi $ we will have 
\begin{equation}  \label{2.16}
^c\nabla _u\xi =\nabla _u\xi -[\overline{g}(^F\omega )(\xi )+l.\xi -\frac 12.%
\overline{g}(\nabla _ug)(\xi )]\text{ ,}
\end{equation}

where $l=g(\xi ,u)$, $\overline{g}(g)(\xi )=\overline{g}[g(\xi )]=\xi $.
Then $^c\nabla _u\xi =0$ is equivalent to a conformal transport in the form 
\begin{equation}  \label{2.17}
\nabla _u\xi =[\overline{g}(^F\omega )(\xi )+l.\xi -\frac 12.\overline{g}%
(\nabla _ug)(\xi )]\text{ .}
\end{equation}

It is obvious that if the vector field $\xi $ is orthogonal to $u$ $%
[l=g(u,\xi )=0]$, then $\nabla _u\xi =\overline{g}[^F\omega (\xi )]-\frac 12.%
\overline{g}[(\nabla _ug)(\xi )]$ is a generalized Fermi-Walker transport of
type $C$ along a non-null vector field $u$.

\textit{Remark}. If $q=\xi $ (or $b=\xi $, or $b=q=\xi $) in the expression
for $^c\nabla _u\xi $, then $^c\nabla _u\xi $ is not more a linear transport
with respect to the contravariant vector field $\xi $.

(b) If we chose $b=q=\xi $ in the expression for $^c\nabla _u\xi $, it
follows that 
\begin{equation}  \label{2.18}
^c\nabla _u\xi =\nabla _u\xi -\{l.[\overline{g}(^F\omega )(\xi )+\xi ]-\frac
12.\overline{g}(\nabla _ug)(\xi )\}\text{ .}
\end{equation}

Then $^c\nabla _u\xi =0$ is equivalent to a conformal transport in the form

\begin{equation}  \label{2.19}
\nabla _u\xi =l.[\overline{g}(^F\omega )(\xi )+\xi ]-\frac 12.\overline{g}%
(\nabla _ug)(\xi )\text{ .}
\end{equation}

For an orthogonal to $u$ vector field $\xi $ $[l=g(\xi ,u)=0]$ we obtain a
Fermi-Walker transport of type $B$ for the vector field $\xi $%
\begin{equation}  \label{2.20}
\nabla _u\xi =-\frac 12.\overline{g}(\nabla _ug)(\xi )=\overline{g}[(\nabla
_uB)(\xi )]
\end{equation}

\noindent with $B=-\frac 12.g$.

In this case the condition $e=g(u,u)\neq 0$ is not used and $u$ could be a
non-null contravariant vector field ($e\neq 0$) as well as a null
contravariant vector field ($e=0$).

(c) One of the vector fields $b$ and $q$ or both vectors could also be
related to the Weyl's covariant vector in a $(L_n,g)$-space if we represent $%
\nabla _ug$ by means of its trace-free part and its trace part in the form 
\begin{equation}
\nabla _ug=\,^s\nabla _ug+\frac 1n.Q_u.g\text{ , \thinspace \thinspace
\thinspace \thinspace \thinspace \thinspace \thinspace \thinspace \thinspace
\thinspace \thinspace \thinspace }\dim M=n\text{ ,}  \label{2.21}
\end{equation}

\noindent where $\overline{g}[^s\nabla _ug]=0$, $Q_u=\overline{g}[\nabla _ug%
]=g^{kl}.g_{kl;j}.u^j=Q_j.u^j$, \thinspace $Q_j=g^{kl}.g_{kl;j}$.

\textit{Remark}. The covariant vector $\overline{Q}=\frac 1n.Q=\frac
1n.Q_j.dx^j $ is called \textit{Weyl's covector}. The operator $^s\nabla
_u=\nabla _u-\frac 1n.Q_u$ is called \textit{trace-free covariant operator}.

(c$_1$) If we chose $b=\frac 1e.u$ and $q=\overline{g}(\overline{Q})=%
\widetilde{Q}$, then 
\begin{equation}  \label{2.22}
^c\nabla _u=\nabla _u-[\overline{g}(^F\omega )+g(\widetilde{Q},u).\overline{g%
}(g)-\frac 12.\overline{g}(\nabla _ug)]\text{ ,}
\end{equation}

\noindent and $^c\nabla _u\xi =0$ will have the form 
\begin{equation}
\nabla _u\xi =\overline{g}[(^F\omega )(\xi )]+g(\widetilde{Q},u).\xi -\frac
12.\overline{g}[(\nabla _ug)(\xi )]\text{ .}  \label{2.23}
\end{equation}

If the vector field $u$ is orthogonal to $\widetilde{Q}$, i. e. $g(%
\widetilde{Q},u)=0$, then $\nabla _u\xi =\overline{g}[(^F\omega )(\xi )%
]-\frac 12.\overline{g}[(\nabla _ug)(\xi )]$ and we have a generalized
Fermi-Walker transport of type $C$ along a non-null vector field $u$.

(c$_2$) If we chose $b=q=\overline{Q}$ in the expression for $^c\nabla _u$,
we will have the relation 
\begin{equation}  \label{2.24}
^c\nabla _u\xi =\nabla _u\xi -\{g(\widetilde{Q},u).[\overline{g}(^F\omega
)(\xi )+\xi ]-\frac 12.\overline{g}(\nabla _ug)(\xi )\}\text{ .}
\end{equation}

Then $^c\nabla _u\xi =0$ is equivalent to 
\begin{equation}  \label{2.25}
\nabla _u\xi =g(\widetilde{Q},u).[\overline{g}(^F\omega )(\xi )+\xi ]-\frac
12.\overline{g}(\nabla _ug)(\xi )\text{ . }
\end{equation}

For an orthogonal to $\widetilde{Q}$ vector field $u$ [$g(\widetilde{Q},u)=0$%
] we obtain a Fermi-Walker transport of type $B$ for the vector field $\xi
:\nabla _u\xi =-\frac 12.\overline{g}[(\nabla _ug)(\xi )]$ as in the case
(b) for $l=0$. Here $u$ could also be a non-null ($e\neq 0$) or null ($e=0$)
contravariant vector field.

\textit{Remark}. The assumption $g(\widetilde{Q},u)=0$ contradicts to the
physical interpretation of $\widetilde{Q}$ as a vector potential $A$ of the
electromagnetic field in $W_n$-spaces ($n=4$) \cite{Landau} because of the
relation $g(A,u)=e_0.[g(u,u)]/[g(R,u)\neq 0$. In our case here $\widetilde{Q}
$ could be more related to the Lorentz force $_LF=\overline{g}[F(u)]$ than
with $A$. The tensor $F$ is the electromagnetic tensor: $%
F=dA=(A_{j,i}-A_{i,j}).dx^i\wedge dx^j$.

\textit{Remark}. In $(\overline{L}_n,g)$-spaces \cite{Manoff-1} analogous
considerations could also be done.

\section{Conformal transports for null vector fields}

$l_\xi ^2=0$ is fulfilled for a null contravariant vector field $\xi $. Then
the condition for a conformal transport of $\xi $%
\begin{equation}  \label{3.1}
l_\xi .ul_\xi =\pm \frac 12.[(\nabla _ug)(\xi ,\xi )+2.g(\overline{A}_u\xi
,\xi )]=f(u).l_\xi ^2\text{ }
\end{equation}

\noindent with $\overline{A}_u=\overline{g}(C(u))=g(b,u).\overline{g}%
(^F\omega )+g(q,u).\overline{g}(g)-\frac 12.\overline{g}(\nabla _ug)$ is
fulfilled identically for $\forall f(u)\in C^r(M)$. For two null
contravariant vector fields $\xi $ and $\eta $ [$l_\xi ^2=0$, $l_\eta ^2=0$]
under a conformal transport the relation 
\begin{equation*}
l_\xi .l_\eta .u[\cos (\xi ,\eta )]=(\nabla _ug)(\xi ,\eta )+g(\overline{A}%
_u\xi ,\eta )+g(\xi ,\overline{A}_u\eta )=0\text{ } 
\end{equation*}

\noindent is also identically fulfilled.

\textit{Remark}. Contravariant null vector fields fulfil also identically
the conditions for Fermi-Walker transports.

Therefore, \textit{every contravariant null vector field }$\xi $\textit{\
obeys automatically the conditions for a conformal transport. }Moreover,%
\textit{\ the (right) angle between two contravariant null vector fields }$%
\xi $\textit{\ and }$\eta $\textit{\ is automatically preserved under a
conformal transport.}

If the contravariant vector field $u$ is a null vector field [$e=g(u,u)=0$]
we cannot choose $b$ as $b=\frac 1e.u$ but we can consider it as $b=u$. In
such a case $g(b,u)=g(u,u)=0$ and $\overline{A}_u$ will have the form 
\begin{equation}  \label{3.2}
\overline{A}_u=g(q,u).\overline{g}(g)-\frac 12.\overline{g}(g)=\overline{g}%
(C(u))\text{ .}
\end{equation}

If we define further $q$ as a contravariant null vector field [$g(q,q)=0$],
then $q$ could be written in the form $q=\alpha .u$ [$\alpha \in C^r(M)$]
and $\overline{A}_u$ will have the form 
\begin{equation}  \label{3.3}
\overline{A}_u=-\frac 12.\overline{g}(\nabla _ug)\text{ }
\end{equation}

\noindent identical with the form of $\overline{A}_u$ for a Fermi-Walker
transport of type $B$.

\section{The length of a contravariant vector field as harmonic oscillator
over $(L_n,g)$-spaces}

Let us now try to find out a solution of the following problem: under which
conditions the length $l_\xi $ of a non-null vector field $\xi $ moving
under conformal transport along a contravariant vector field $u$ could
fulfil the equation of a harmonic oscillator in the form 
\begin{equation}  \label{4.1}
\frac{d^2l_\xi }{d\tau ^2}+\omega _0^2.l_\xi =0\text{ , \thinspace
\thinspace \thinspace \thinspace \thinspace \thinspace \thinspace \thinspace
\thinspace \thinspace }\omega _0^2=\text{ const. }\geq 0\text{ ,\thinspace
\thinspace \thinspace \thinspace \thinspace \thinspace \thinspace \thinspace
\thinspace \thinspace \thinspace }u^i=\frac{dx^i}{d\tau }\text{ , }%
\,\,\,\,\,\,u=\frac d{d\tau }\text{ . \thinspace \thinspace }
\end{equation}

Lets a contravariant non-null vector field $\xi $ be given, moving under
conformal transport along a trajectory $x^k(\tau )$ with the tangential
vector $u=\frac d{d\tau }$ in a $(L_n,g)$-space. The change of the length of 
$\xi $ along the trajectory is given as [see (\ref{2.1}) and (\ref{2.14})] 
\begin{equation}  \label{4.2}
\frac{dl_\xi }{d\tau }=g(q,u).l_\xi \text{ , \thinspace \thinspace
\thinspace \thinspace \thinspace \thinspace \thinspace \thinspace \thinspace
\thinspace \thinspace \thinspace \thinspace }\xi ,q,u\in T(M)\text{ .}
\end{equation}

Since $q$ is an arbitrary given contravariant vector field we can specify
its structure in a way allowing us to consider $g(q,u)$ as an invariant
scalar function depending only of the parameter $\tau $ of the trajectory $%
x^k(\tau )$. A possible choice of $q$ fulfilling this precondition is 
\begin{equation}  \label{4.3}
q=\omega (\tau ).\frac 1e.u\text{ , \thinspace \thinspace \thinspace
\thinspace \thinspace \thinspace \thinspace \thinspace \thinspace \thinspace
\thinspace }\omega (\tau )=\omega (x^k(\tau ))\in C^r(M)\text{ , \thinspace
\thinspace \thinspace \thinspace }r\geq 1\text{, \thinspace \thinspace
\thinspace \thinspace }e=g(u,u)\neq 0\text{ . }
\end{equation}

Then we have 
\begin{equation}  \label{4.4}
\frac{dl_\xi }{d\tau }=\omega (\tau ).l_\xi
\end{equation}

\textit{Remark}. The solution of this equation for $l_\xi $ can be found in
the form 
\begin{equation}  \label{4.5}
l_\xi =l_{\xi 0}.\exp (\int \omega (\tau ).d\tau )\text{ , \thinspace
\thinspace \thinspace \thinspace \thinspace \thinspace \thinspace \thinspace
\thinspace }l_{\xi 0}=\text{ const., \thinspace \thinspace \thinspace }l_\xi
>0\text{ ,}
\end{equation}

The equation for $d^2l_\xi /d\tau ^2$ could be written in the form 
\begin{equation}
\frac{d^2l_\xi }{d\tau ^2}-\left\{ \frac{d\omega (\tau )}{d\tau }+[\omega
(\tau )]^2\right\} .l_\xi =0\text{ .}  \label{4.7}
\end{equation}

It could have the form of an oscillator equation if the arbitrary given
until now function $w(\tau )$ fulfils the equation 
\begin{equation}  \label{4.8}
\frac{d\omega (\tau )}{d\tau }+[\omega (\tau )]^2=-\,\,\omega _0^2\leq 0%
\text{ }
\end{equation}

and could be determined by means of this (additional) condition. The last
equation is a special case of Riccati equation and at the same time an
equation with separable variables \cite{Kamke}. By the use of the
substitution $u^{\prime }=\omega .u$, $u^{\prime }=du/d\tau $, it could be
written in the form of a homogeneous harmonic oscillator equation for $u$%
\begin{equation}  \label{4.9}
u^{\prime \prime }+\omega _0^2.u=0\text{ ,}
\end{equation}

where 
\begin{equation*}
\omega (\tau )=\frac{u^{\prime }(\tau )}{u(\tau )}\text{ , \thinspace
\thinspace \thinspace \thinspace \thinspace \thinspace \thinspace }\omega
^{\prime }=-\frac{(u^{\prime })^2}{u^2}+\frac{u^{\prime \prime }}u\text{ ,
\thinspace \thinspace \thinspace \thinspace \thinspace }\omega ^{\prime
}+\omega ^2=-\omega _0^2\text{ .} 
\end{equation*}

For $\omega _0^2\geq 0$ (\ref{4.9}) has the solutions:

(a) $\omega _0^2=0:u^{\prime }=C_2.\tau $ , \thinspace \thinspace \thinspace 
$u=C_1+C_2.\tau $ , $C_1,C_2=$ const.

(b) $\omega _0^2>0:u=C_1.\cos \omega _0.\tau +C_2.\sin \omega _0.\tau $ , $%
C_1,C_2=$ const.

The solution for $\omega $ will have the forms:

(a) $\omega _0^2=0$: 
\begin{equation}  \label{4.10}
\omega (\tau )=\frac{C_2.\tau }{C_1+C_2.\tau }=\frac \tau {a+\tau }\text{ ,
\thinspace \thinspace \thinspace \thinspace \thinspace \thinspace }a=\frac{%
C_1}{C_2}=\text{const., \thinspace \thinspace \thinspace \thinspace
\thinspace }C_2\neq 0\text{ .}
\end{equation}

(b) $\omega _0^2>0$: 
\begin{equation}  \label{4.11}
\omega (\tau )=\frac{u^{\prime }}u=\omega _0.\frac{C_2.\cos \omega _0.\tau
-C_1.\sin \omega _0.\tau }{C_1.\cos \omega _0.\tau +C_2.\sin \omega _0.\tau }%
\text{ , \thinspace \thinspace }
\end{equation}
\begin{equation}  \label{4.12}
\omega (\tau )=\omega _0.\frac{1-a.\tan \omega _0.\tau }{a+\tan \omega
_0.\tau }\text{ , \thinspace \thinspace \thinspace \thinspace \thinspace
\thinspace \thinspace \thinspace }a=\frac{C_1}{C_2}=\text{ const. ,
\thinspace \thinspace \thinspace \thinspace \thinspace \thinspace \thinspace
\thinspace \thinspace }C_2\neq 0\text{ .}
\end{equation}

(b$_1$) For $C_2=0:\omega (\tau )=-\omega _0.\tan \omega _0.\tau $ .

(b$_2$) For $C_1=0:\omega (\tau )=\omega _0.\cot \omega _0.\tau =\omega
_0/\tan \omega _0.\tau $.

The integral curve through a point $(\tau _1,\omega (\tau _1))$ can be
written in the forms \cite{Kamke} 
\begin{equation}
\omega (\tau )= 
\begin{array}{cc}
\frac{\omega (\tau _1)}{1+\omega (\tau _1).(\tau -\tau _1)} & 
\begin{array}{c}
\text{for \thinspace \thinspace \thinspace }\omega _0^2=0\text{ ,}
\end{array}
\\ 
\omega _0.\frac{\omega (\tau _1)-\omega _0.\tan \omega _0(\tau -\tau _1)}{%
\omega _0+\omega (\tau _1).\tan \omega _0(\tau -\tau _1)} & 
\begin{array}{c}
\text{ for }\omega _0^2>0\text{ .}
\end{array}
\end{array}
\label{4.12a}
\end{equation}

Therefore, if $l_\xi $ should be a harmonic oscillator in a $(L_n,g)$-space
with a given constant frequency $\omega _0$ the function $\omega (\tau )$
has to obey a Riccati equation determining its structure. In this case, the
length $l_\xi $ will be a homogeneous harmonic oscillator obeying at the
same time the equation 
\begin{equation}  \label{4.13}
\frac{dl_\xi }{d\tau }=\omega _0.\frac{1-a.\tan \omega _0.\tau }{a+\tan
\omega _0.\tau }.l_\xi \text{ , \thinspace \thinspace \thinspace \thinspace
\thinspace \thinspace \thinspace \thinspace \thinspace \thinspace \thinspace
\thinspace \thinspace \thinspace }a=\frac{C_1}{C_2}=\text{ const. ,}
\end{equation}

\noindent which appears in general as a solution of the homogeneous harmonic
oscillator's equation 
\begin{equation*}
\frac{d^2l_\xi }{d\tau ^2}+\omega _0^2.l_\xi =0\text{ , \thinspace
\thinspace \thinspace \thinspace \thinspace \thinspace \thinspace \thinspace
\thinspace \thinspace }\omega _0^2=\text{ const. }\geq 0\text{ .} 
\end{equation*}

If we could simulate by the use of an appropriate experimental device the
change $(dl_\xi /d\tau )$ as given in its equation (\ref{4.13}), then we can
be sure that $l_\xi $ will swing as a harmonic oscillator with the frequency 
$\omega _0$ over a $(L_n.g)$-space. This could allow us to investigate
experimentally the influence of physical interactions on the length $l_\xi $
of a vector field $\xi $ moving under a conformal transport over a $(L_n,g)$%
-space. On the other hand, if we can register (as an observer) from our
basic trajectory a change of $l_\xi $ in accordance with the equation for $%
(dl_\xi /d\tau )$, then we can conclude that $l_\xi $ moves as harmonic
oscillator under the external (or internal) forces. Since the considered
here problem is related to the kinetic aspect of the motion of $l_\xi $, the
dynamic aspect should be introduced by means of a concrete field
(gravitational) theory, which is not a subject of the above considerations.

\subsection{Geometrical and physical interpretation of the function $\protect%
\omega (\protect\tau )$}

The covariant derivative $\nabla _u\xi $ of a contravariant vector field $%
\xi $ along a contravariant vector field $u$ in a $(L_n,g)$-space can be
represented in the form \cite{Manoff-5} 
\begin{equation}  \label{4.14}
\nabla _u\xi =\frac{\overline{l}}e.u+\overline{g}[h_u(\nabla _u\xi )]=\frac{%
\overline{l}}e.u+\,_{rel}v\text{ , \thinspace \thinspace \thinspace
\thinspace \thinspace \thinspace \thinspace \thinspace \thinspace \thinspace
\thinspace \thinspace \thinspace }\overline{l}=g(\nabla _u\xi ,u)
\end{equation}

where 
\begin{equation}  \label{4.15}
_{rel}v=\overline{g}[h_u(\nabla _u\xi )]=\overline{g}(h_u)(\frac
le.a-\pounds _\xi u)+\overline{g}[d(\xi )]\text{ , \thinspace \thinspace
\thinspace }h_u=g-\frac 1e.g(u)\otimes g(u)\text{ ,}
\end{equation}
\begin{equation}  \label{4.16}
a=\nabla _uu\text{ , \thinspace \thinspace \thinspace \thinspace \thinspace
\thinspace \thinspace \thinspace }d=\sigma +\omega +\frac 1{n-1}.\theta .h_u%
\text{ , \thinspace \thinspace }g(\nabla _u\xi ,u)=ul-(\nabla _ug)(\xi
,u)-g(\xi ,a)\text{ .}
\end{equation}

The tensor $\sigma $ is the shear velocity tensor (shear), $\omega $ is the
rotation velocity tensor (rotation), the invariant $\theta $ is the
expansion velocity invariant (expansion), the tensor $d$ is the deformation
velocity tensor (deformation), the invariant $l=g(\xi ,u)$, the vector field 
$_{rel}v$ is the relative velocity. If $\xi $ is an orthogonal to $u$ vector
field [$\xi =\xi _{\perp }=\overline{g}[h_u(\xi )]$], then $l=0$ and under
the additional precondition $\pounds _\xi u=-\pounds _u\xi =0$ the
expression for $_{rel}v$ will take the form 
\begin{equation}
_{rel}v=\overline{g}[d(\xi _{\perp })]=\overline{g}[\sigma (\xi _{\perp })]+%
\overline{g}[\omega (\xi _{\perp })]+\frac 1{n-1}.\theta .\xi _{\perp }\text{
.}  \label{4.17}
\end{equation}

The rate of change of the length $l_{\xi _{\perp }}$ of the vector field $%
\xi _{\perp }$ (along the vector field $u=\frac d{d\tau }$) in $U_n$- or $%
V_n $-spaces [$\nabla _ug=0$ for $\forall u\in T(M)$], under the conditions $%
l=0$ [$\xi =\xi _{\perp }=\overline{g}[h_u(\xi )]$] and $\pounds _u\xi =0$,
can be found in the form 
\begin{equation}  \label{4.18}
ul_{\xi _{\perp }}=\frac{dl_{\xi _{\perp }}}{d\tau }=\pm \frac 1{l_{\xi
_{\perp }}}.d(\xi _{\perp },\xi _{\perp })=\pm \frac 1{l_{\xi _{\perp
}}}.\sigma (\xi _{\perp },\xi _{\perp })+\frac 1{n-1}.\theta .l_{\xi _{\perp
}}\text{ ,\thinspace \thinspace \thinspace \thinspace \thinspace \thinspace
\thinspace \thinspace \thinspace \thinspace \thinspace \thinspace \thinspace
\thinspace }l_{\xi _{\perp }}\neq 0\text{ .}
\end{equation}

\textit{Remark}. The sign $\pm $ depends on the sign of the metric $g$ (for $%
n=4 $, sign $g=\pm 2$).

If a $U_n$- or a $V_n$-space admits a shear-free non-null auto-parallel
vector field $u$ ($\nabla _uu=a=0$), then $\sigma =0$ and 
\begin{equation}  \label{4.19}
\nabla _u\xi _{\perp }=\,_{rel}v=\overline{g}[\omega (\xi _{\perp })]+\frac
1{n-1}.\theta .\xi _{\perp }=\overline{g}[\omega (\xi _{\perp })]+\frac
1{n-1}.\theta .\overline{g}[g(\xi _{\perp })]=
\end{equation}

\begin{equation}  \label{4.20}
=\overline{g}[C(u)(\xi _{\perp })]=\overline{A}_u\xi _{\perp }\text{
,\thinspace \thinspace \thinspace \thinspace \thinspace \thinspace
\thinspace \thinspace \thinspace \thinspace \thinspace \thinspace }%
C(u)=\omega +\frac 1{n-1}.\theta .g\text{ .}
\end{equation}
\begin{equation}  \label{4.21}
\frac{dl_{\xi _{\perp }}}{d\tau }=\frac 1{n-1}.\theta .l_{\xi _{\perp }}%
\text{ .}
\end{equation}

Therefore, the vector field $\xi _{\perp }$ undergoes a conformal transport
along $u$. A comparison with (\ref{4.4}) show us that in this case we can
choose the arbitrary given function $\omega (\tau )$ as 
\begin{equation}  \label{4.22}
\omega (\tau )=\frac 1{n-1}.\theta \text{ .}
\end{equation}

The last fact leads to the conclusion that $\omega (\tau )$ could be related
to the expansion velocity $\theta $ in a $U_n$- or a $V_n$-space. In $%
(L_n,g) $-spaces it could preserve its interpretation.

\section{Conclusions}

In the present paper we have considered types of transports more general
than the Fermi-Walker transports. They are called conformal transports over $%
(L_n,g)$-spaces. In an analogous way as in the case of Fermi-Walker
transports a conformal covariant differential operator and its corresponding
conformal derivative are determined and discussed over $(L_n,g)$-spaces.
Different special types of conformal transports are considered inducing also
Fermi-Walker transports for orthogonal vector fields as special cases.
Conditions under which the length of a non-null contravariant vector field
will swing as a homogeneous harmonic oscillator with given frequency are
established. The results obtained regardless of any concrete field
(gravitational) theory could have direct applications in such types of
theories.

\begin{center}
\textsc{Acknowledgments}
\end{center}

This work is supported in part by the National Science Foundation of
Bulgaria under Grant No. F-642.

\end{document}